\documentclass [floatfix, showpacs, preprint, showkeys, aps, pra]{revtex4-1}

\usepackage{amsmath}
\usepackage{graphicx}
\usepackage{bbm}
\usepackage{amssymb}
\usepackage[english]{babel}
\usepackage[latin1]{inputenc}
\usepackage{lmodern}
\usepackage[T1]{fontenc}

\DeclareMathOperator{\e}{{\displaystyle e}}
\DeclareMathOperator{\de}{{\displaystyle d}}

\graphicspath{{Figs/}}

\newcommand{\TreJ}[6]{
  \begin{pmatrix}
    #1 & #2 & #3 \\
    #4 & #5 & #6
  \end{pmatrix}
}

\newcommand{\SeiJ}[6]{
  \left\{
    \begin{matrix}
      #1 & #2 & #3 \\
      #4 & #5 & #6
    \end{matrix}
  \right\}
}

\begin{document}
\title{Atomic orientation 
by a 
broadly frequency-modulated radiation: theory and experiment.}
\author{G. Bevilacqua}
\author{V. Biancalana}
\affiliation {DIISM, Universit\`a di Siena, via Roma 56, 53100 Siena, Italy.}

\author{Y. Dancheva}
\affiliation {Universit\`a di Siena,  via Roma 56, 53100 Siena, Italy.}

\begin{abstract}
  We investigate magnetic resonances  driven in thermal vapour of alkali atoms 
  by laser  radiation broadly modulated at a  frequency resonant with
  the  Zeeman splitting.  A  model accounting  for both  hyperfine and
  Zeeman  pumping   is  developed   and  its  results are compared  with
  experimental   measurements  performed   at  relatively   weak  pump
  irradiance.   The  interplay   between  the  two  pumping  processes
  generates  intriguing interaction  conditions,  often overlooked  by
  simplified models.
\end{abstract}

\date{\today}



\maketitle

\section{Introduction} 
\label{sec:introduction}


Optical pumping processes in atomic samples \cite{happer_book_10} have
been subject  of intensive theoretical and  experimental studies since
the  60s   \cite{happer_rmp_72},  and   have  been  used   in  several
applications   including,    laser   cooling   \cite{metcalf_book_99},
molecular spectroscopy \cite{auzinsh_book_05} and atomic magnetometry.
Atomic magnetometers are nowadays available as commercial devices, but
further  research is  presently  carried out  
to optimize the performance, as  well as to
better understand  phenomena and mechanisms  which subtly act  in this
kind of apparatuses.

The interest in precise and sensitive magnetic field measurements led
to  a  revival  of the research  in  magnetometry,  and  particularly  in
the optical-atomic sensors.   Optical magnetometers were  recently subject
to impressive  advances in terms  of sensitivity.  The  possibility of
absolute  field  measurements,  the  low  operation  costs  and  power
consumption,  the robustness, and  the potential  for miniaturization,
let  these devices compete  with superconducting  quantum interference
devices,  traditionally  regarded  as  state-of-the-art  magnetometric
sensors.

The  typical  working  principle  of an  atomic  optical  magnetometer
\cite{aleksandrov_pu_09}  is based  on  the preparation  of an  atomic
state using optical pumping and on the detection of its time-evolution
driven by the magnetic field  under measurement.  Some recent works on
atomic magnetometry  have addressed time-domain  operation techniques,
where the atomic  state is first prepared and then  is followed in its
free evolution within the decay time \cite{lenci_pra_14}. In contrast,
most  of the  approaches reported  in the  literature are  based  on a
frequency-domain  detection  \cite{BudkerRomalis_nature2007}. In  this
case,  a steady-state  condition is  reached, by  means of  a periodic
regeneration of the atomic state  to be analyzed. This regeneration is
obtained  by applying  an  appropriate optical  radiation having  some
parameter periodically modulated in resonance (or near-resonance) with
the  evolution of the  atomic state.   Experiments have  been reported
where the modulated  parameter of the pump radiation  is its amplitude
\cite{suter91,  schultze_oe_12,  rosatzin_josab_90}, its  polarization
\cite{klepel_oc_92, breschi_pra_13,  breschi_apl_14,bevilacqua:2014:hv},   
or   its   optical   frequency
\cite{belfi_josab_07, acosta_pra_06}. Different macroscopic quantities
have  been chosen  to  be measured  as  well, such  as the  absorption
\cite{breschi_pra_13},    the   polarization    rotation   \cite{budker_10,
  bevilacqua_pra_12,  romalis_13},  or   (in  similar  experiments
based  on solid-state  samples)  the fluorescence  \cite{wolf_prx_15},
all opening  an indirect way to follow the vapour magnetization.

Optical pumping is  often applied in regime of  strong intensity where
power  broadening and  non linear  dependence on  the  laser intensity
occur.  Studies in  low intensity regime are also
reported \cite{sydoryk_pra_08}.

Our study concerns a  setup developed for precise atomic magnetometry,
which here is  operated in a condition of  weak excitation intensity.  The atomic
sample  is illuminated  by  two  collinear laser  beams.   One of  them
(modulated  beam, MB  in  the following)  is  frequency modulated  and
circularly  polarized, and  the  second one  (detection  beam, DB)  is
linearly polarized with polarization plane rotated by the time
dependent circular  birefringence of the sample.  In  other terms, the
MB induces  a magnetic dipole  that precesses at the  Larmor frequency
and the dipole component parallel to the beams is monitored.

The  MB  is broadly  modulated  in  frequency,  thus both  the  ground
hyperfine   states  of  the  atomic  vapour   are  excited  with
non-vanishing rates.  Such broad modulation gives rise to an important
interplay between  hyperfine and Zeeman pumping that takes advantages in optical magnetometry
\cite{bevilacqua_apbarx_16}.
The proposed excitation scheme not  only simplifies the setup (pump-repump scheme
is  often  applied  as  an  alternative), but  has  the  potential  of significantly 
increasing   the  signal  without increasing the   magnetic  resonances width,
particularly  at  higher  intensities,  with obvious practical
implications.

In this  work we  address mainly  the aspects related  to the  wide MB
frequency  modulation, restricting  the investigation  to a  regime of
relatively  weak  intensity, deferring  the  analysis  of the  intense
pumping to another study.
We  develop a  model considering the MB interaction with the whole level structure
of the $D_1$ Cs transitions: a  point which is often overlooked in the
literature. We  obtain a modified  version of the Larmor  equation for
the magnetization created in a given ground state Zeeman multiplet.
An analytical expression for the magnetization amplitude, pointing out
the dependence on the  MB modulation parameters,  is found and
it matches very well with the experiment.


The  paper  is organized  as  follows:  in Section~\ref{sec:setup}  we
briefly      describe     the      experimental      apparatus;     in
Section~\ref{sec:model} the theoretical  model is reported; finally in
Section~\ref{sec:res} we discuss and compare the theoretical and the experimental results.

\section{Experimental setup} 
\label{sec:setup}

\begin{figure}[htbp] \centering
\vspace{12pt}
\includegraphics [width=\columnwidth] {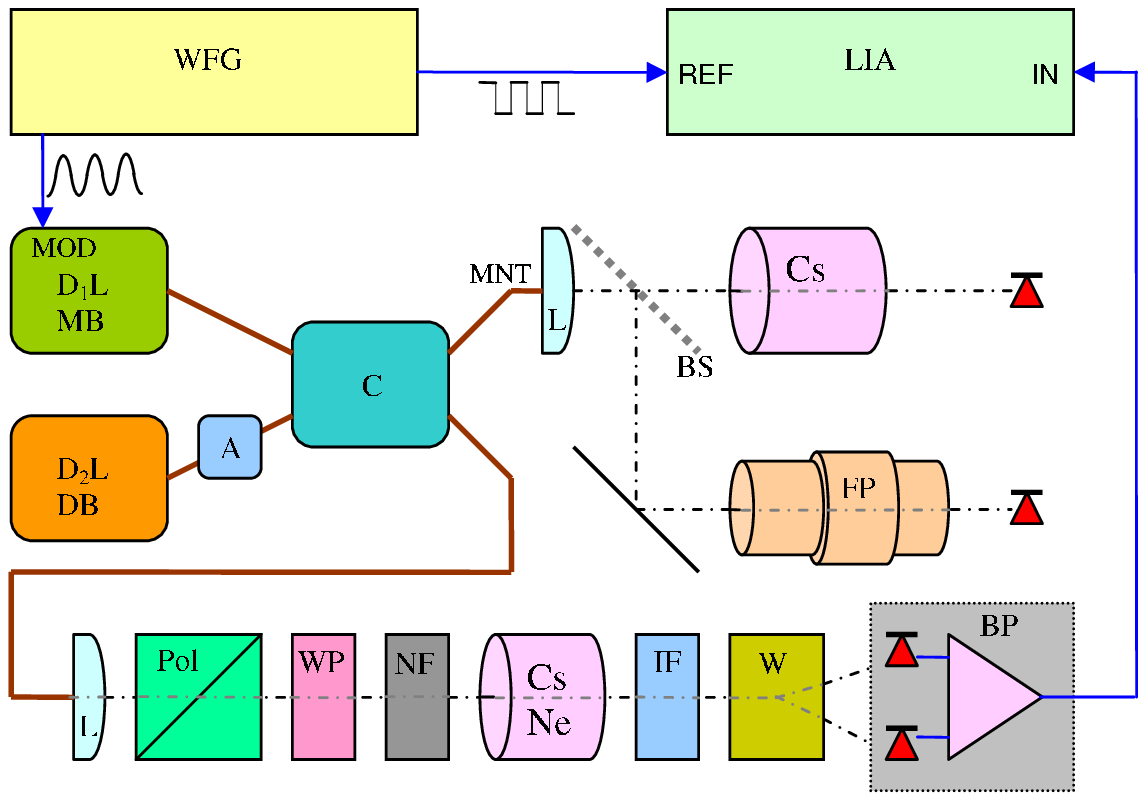}
\caption{Schematics of the  setup. WFG - waveform generator; D$_1$L - pumping laser (MB) at $894$~nm; D$_2$L (DB)
  detection  laser  at   $852$~nm;  A  -  attenuator;  C   -  single  mode,
  polarization maintaining $2\times 2$ fiber  coupler; L - lens, BS - beamsplitter, Pol -
  polarizer;  WP -  multiorder waveplate  acting  as quarter-$\lambda$
  plate for $894$~nm and as  full-$\lambda$ for $852$~nm; NF - neutral
  filter;  Cs-Ne -  Cesium  cell with buffer gas;  IF -  interference  filter  stopping
  $894$~nm; W - Wollaston analizer; BP- balanced polarimeter for 
  polarization  rotation detection, which  includes photo  diodes 
  and differential transimpedance  amplifier; LIA - lockin amplifier. In the monitor (MNT) channel,  Cs - Cesium vacuum cell and FP  -
  Fabry-Perot interferometer are used to monitor the radiations' parameters.}
\label{setup}
\end{figure}

A  detailed  description  of  the  experimental  set-up  is  given  in
Refs.\cite{bevilacqua_pra_12,                     
  bevilacqua_apbarx_16}.  Briefly,  Cs vapour is contained in a sealed cell, where  buffer gas is added to  counteract
time-of-flight  line   broadening  of  the   magnetic  resonances  and
to increase  the optical  pumping  effect.  The Cs Atoms are  optically  pumped by  a
circularly polarized, near resonant  laser (MB) light at 894~nm ($D_1$
Cs line).   The cell  is at room  temperature and in a  highly homogeneous
magnetic field.   A balanced polarimeter enables the  detection of the
atomic precession, which causes the polarization rotation 
of a linearly polarized beam (DB), near resonant with the $F_g$~=~4 to
$F_e$~=~3,~4,~5 group  of transitions belonging to the  Cs $D_2$ line.
The  set-up  contains two  channels  (see  Fig.~\ref{setup}), which  in
magnetometric applications \cite{bevilacqua_jmr_09, bevilacqua_arnmrs_13, bevilacqua_jmr_16}
are used to  reject common-mode magnetic noise and  to measure local
magnetic variations by means of a differential method.  In
the present  work one of the  channels keeps being used  to detect the
atomic spins precession,  while the other one (monitor, MNT) is used for
precise  determination  of the  DB  and  MB  intensities and  absolute
frequencies.  The DB radiation is  attenuated down to 10 nano-Watt and
kept at a constant frequency, blue detuned by about 2~GHz with respect
to the D$_2$ transition set  starting from $F_g=4$.  The MB radiation,
which  in  magnetometric   applications   was  in  the
milli-Watt range,  here is  attenuated down to  100 nano-Watt  and its
optical frequency  is made  time-dependent through a  junction current
modulation  at a  frequency matching  (or ranging  around)  the Larmor
frequency. 
Both  the MB  and DB
have a circular beam spot about $1~\mathrm{cm}^2$ in size.

The optical frequency of the MB  is monitored by the MNT channel, where  the light is sent to a  fixed length Fabry-Perot interferometer and to a secondary Cs  cell without buffer gas.  Both  the absorption and the
interferometric signals are detected  by  photo-detection stages with
a  bandpass largely exceeding  the MB  modulation frequency.   The two
diagnostics provide both a  relative and an absolute measure of
the instantaneous detuning of the MB frequency. The (fixed) DB optical
frequency is monitored as well,  and it is passively stabilized within
 100 MHz.

A sinusoidal signal  modulates the optical frequency of the MB at  the Larmor frequency,
and  references  a  lock-in  amplifier  detecting  the  polarization
rotation of  the DB.  The Cs  cell is  placed  in  a bias
magnetic  field   of  about   $600$~nT  resulting  from   the  partial
compensation of the environmental field.  Such bias field results in a
Cs magnetic resonance  centered at about 2~kHz.  The  amplitude of the
magnetic  resonance  is  registered  for  various  amplitudes  of  the
modulation signal and as a function of the mean MB optical frequency.  To this
aim the MB optical frequency is slowly scanned by adding a ramp to its
modulation signal.

\section{Model}
\label{sec:model}

\begin{figure}[hbtp]
\begin{center}
\includegraphics[width=0.8\columnwidth]{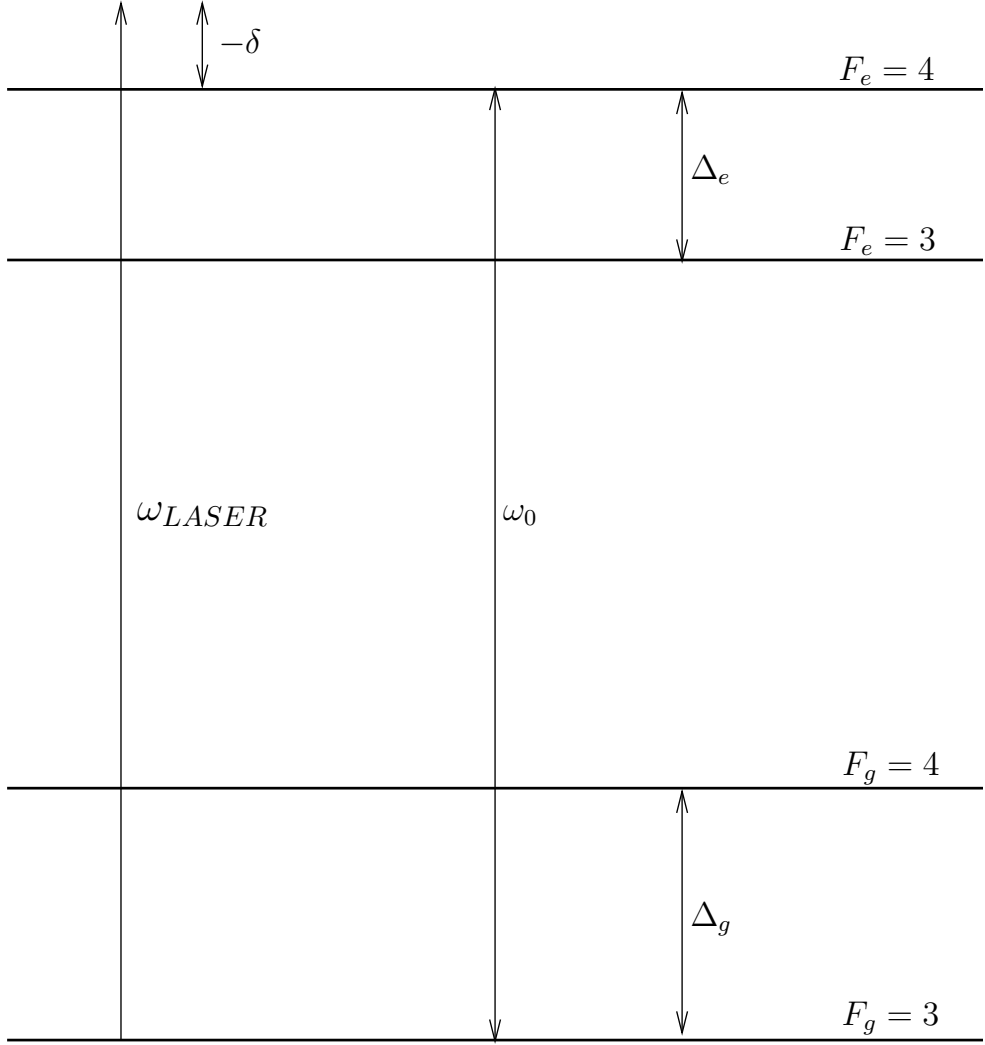}
\end{center}
\caption{Level scheme of Cs $D_1$ line.}
\label{fig:levels}
\end{figure}

To develop  a theoretical model  that describes the time  evolution of
the monitored magnetization, we  consider the whole level structure of
the  $^{133}$Cs $D_1$ line.  With reference  to Fig.~\ref{fig:levels},
the free Hamiltonian in the rotating wave approximation frame reads as
\begin{equation}
  \label{eq:Ho:rwa}
  H_0   =   \Delta_g   \,    \Pi_{g4}   +   \delta   \,   \Pi_{e4}   +
  (\delta-\Delta_e)\, \Pi_{e3} ,
\end{equation}
where  the  projector $\Pi_{g4}$  is  defined  as  $\sum_M |  F_{g}=4,
M\rangle \langle F_{g}=4, M |$. 
Similar expressions hold for the other projectors.

To express the  interaction with the laser field it is  better to adopt a
block-matrix notation 
\begin{equation}
  \label{eq:H:int}
  H_{int} = 
  \begin{pmatrix}
    0 & 0 & W^{\dag}_{e4,g4} & W^{\dag}_{e4,g3} \\
    0 & 0 & W^{\dag}_{e3,g4} & W^{\dag}_{e3,g3} \\
    W_{g4,e4} & W_{g4,e3} & 0 & 0 \\
    W_{g3,e4} & W_{g3,e3} & 0 & 0
  \end{pmatrix},
\end{equation}
where  each   matrix  element  is  a  sub-matrix   defined  using  the
projectors.    For    instance   $W^{\dag}_{e4,g4}   =    -   \Pi_{e4}
\,\mathbf{d}\cdot \boldsymbol{\epsilon} \,\Pi_{g4} \; E_0$, $W_{g4,e4}
= -  \Pi_{g4} \,\mathbf{d}\cdot \boldsymbol{\epsilon}^*  \,\Pi_{e4} \;
E_0$, etc.  
Here  $\boldsymbol{\epsilon}$ is  the  laser polarization
versor,  $E_0$  is the  amplitude  of  the  laser   electric  field  and
$\mathbf{d}$ is the atomic dipole moment.

We need all  these blocks in  our model, because  the laser
modulation can be  very broad and during the  periodic frequency sweep
both the ground states may be resonantly excited. 

The density operator has a similar block-matrix form
\begin{equation}
  \label{eq:def:rho}
  \boldsymbol{\rho} =
  \begin{pmatrix}
    \rho_{e4} & \rho_{e4,e3} & \rho_{e4,g4} & \rho_{e4,g3} \\
    \rho_{e3,e4} & \rho_{e3} & \rho_{e3,g4} & \rho_{e3,g3} \\
    \rho_{g4,e4} & \rho_{g4,e3} & \rho_{g4} & \rho_{g4,g3} \\
    \rho_{g3,e4} & \rho_{g3,e3} & \rho_{g3,g4} & \rho_{g3} 
  \end{pmatrix}.
\end{equation}
The blocks are defined in the manner described above. 
The diagonal  blocks $\rho_{e4}, \rho_{e3}, \rho_{g4}$, and
$\rho_{g3}$  contain  both  the   level  populations  and  the  Zeeman
coherences. The blocks $\rho_{e4,e3} = \rho_{e3,e4}^{\dag}$ and 
$\rho_{g4,g3}   =   \rho_{g3,g4}^{\dag}$   represent   the   hyperfine
coherences, while the remaining blocks represent the optical coherences. 

We  assume that  the hyperfine  coherences can  be  neglected (secular
approximation) and with standard methods we write the Bloch equation:
\begin{equation}
  \label{eq:bloch}
  \boldsymbol{\dot{\rho}} =  -i [ H_0 +  H_{int}, \boldsymbol{\rho}] +
  \mathcal{L}_D\,  \boldsymbol{\rho},
\end{equation}
where the  Liouvillian $\mathcal{L}_D$ takes into  account the effects
of relaxation processes like spontaneous emission and/or collisions. 

As the magnetization is monitored by the DB tuned in the vicinity of the
$F_g=4  \rightarrow J_e=3/2$ transition,  the signal  is substantially 
given by the $| F_{g}=4 \rangle$  state.  We assume that the effect of
the DB  is very weak  and its contribution  to the Hamiltonian  can be
neglected. Hence the Bloch equation \eqref{eq:bloch} contains only the
MB interaction.  To some extent, this approximation  is relaxed in the following (see
Appendix  \ref{sec:der:pumping}). 

After  some   algebra  and  introducing   the  irreducible  components
\cite{omont_book_77, happer_rmp_72}
\begin{align}
  \label{eq:irrep:comp}
  \rho_{g4} & = \sum_{k=0}^{2\,F_{g4}}\,\sum_{q=-k}^k \; m_{k,q} \; T_{k,q}(g4) 
\end{align}
in the hypothesis of weak laser power regime, we find the  final equation for the
ground state $F_g=4$ orientation:
\begin{equation}
  \label{eq:fin:magn}
  \begin{split}
  \dot{\mathbf{x}} = & i \frac{ \omega_L }{\sqrt{2}}
  \begin{pmatrix}
    - \sqrt{2} (\cos \theta + \gamma) & \sin \theta \e^{i\phi} & 0 \\
    \sin   \theta    \e^{-i\phi}   &   -\sqrt{2}\gamma    &   \sin   \theta
    \e^{i\phi} \\
    0 & \sin \theta \e^{-i\phi} & \sqrt{2} (\cos \theta - \gamma)
  \end{pmatrix}\; \mathbf{x} \\
  & + P(t) \begin{pmatrix}
    0\\ 1 \\ 0
  \end{pmatrix} \\
  = & A\, \mathbf{x} 
  + P(t) \mathbf{w},
\end{split}
\end{equation}
where the vector $\mathbf{x}$ is defined as $ \mathbf{x} = ( m_{1,-1},
m_{1,0}, m_{1,1} ) $.  
%

The model produces equations  for both the magnetization (orientation)
and the alignment, however in this work we discuss only  the dynamics of the
orientation.

The pumping rate  $P(t)$ is reported in the Appendix  with full derivation
details.
Notice that Eq.~\eqref{eq:fin:magn} is
essentially  equivalent  to the  Larmor  equation  with an  additional
forcing term, being $M_x \propto (m_{1,1} - m_{1,-1})$, 
$M_y \propto i(m_{1,1} + m_{1,-1})$ and $M_z \propto m_{1,0}$.

\begin{figure}[hbtp]
\begin{center}
\includegraphics[width=\columnwidth]{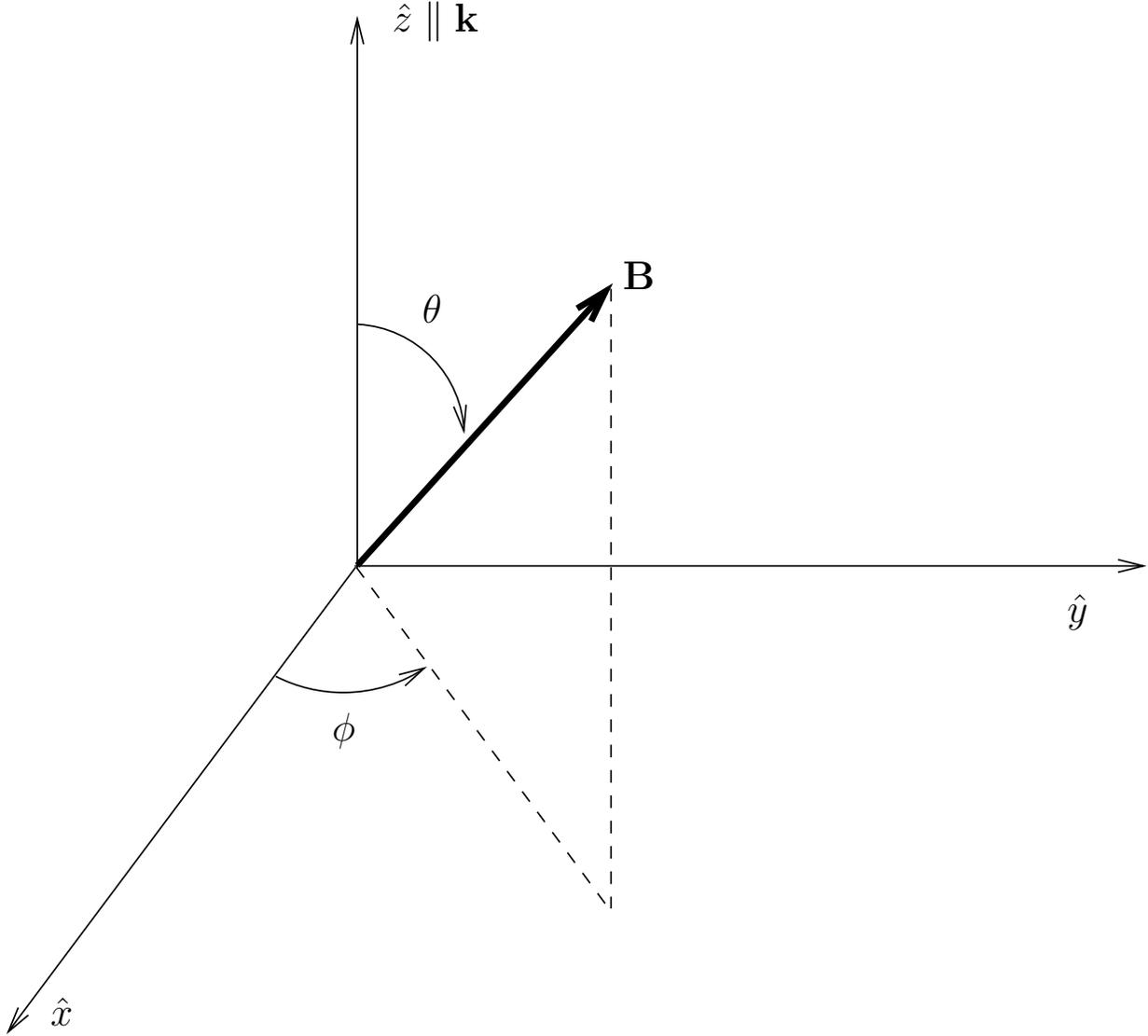}
\end{center}
\caption{Schematic of the geometry used in the model.}
\label{fig:geometry}
\end{figure}

The  Larmor  frequency  is  $\omega_L  =  g_{Fg} \mu_B  B  $.  In  our
experiment $\omega_L$ is in the  kHz range, while the relaxation rates
(longitudinal   and    transverse)   are   in   Hz    range,   so   in
Eq.~\eqref{eq:fin:magn} we used a  single rate $\gamma$.  The geometry
considered in the model is sketched Fig.~\ref{fig:geometry}.

The matrix of coefficients in Eq.~\eqref{eq:fin:magn} can be diagonalized
by a Wigner rotation \cite{sakurai} matrix $U$ so that 
\begin{equation}
  \label{eq:diag}
  U^{\dag} \,A\, U = A_D = 
  \begin{pmatrix} 
    -i \omega_L - \gamma & 0 & 0 \\
    0 & -\gamma & 0 \\
    0 & 0 & i\omega_L - \gamma 
    \end{pmatrix}
\end{equation}
and the full solution is 
\begin{equation}
  \label{eq:full:sol:x}
  \mathbf{x}(t) = U\e^{A_D\, t}U^{\dag}\,\mathbf{x}(0) + 
  U \int_0^t \e^{A_D (t  - t^{\prime})} P(t^{\prime})  \de t^{\prime} \;
  U^{\dag} \mathbf{w}. 
\end{equation}
After a time interval much longer than $ 1/\gamma $, the free solution
fades  away and  the last  term sets  as the  steady-state orientation
$\mathbf{x}_{SS}$.
Introducing the Fourier components of the pumping term 
\begin{equation}
  \label{eq:fourier:pumping}
  P(t) = \sum_{n=-\infty}^{+\infty} P_n \e^{i\,n\,\Omega\,t}, 
\end{equation}
where $\Omega \approx \omega_L $ is the modulation frequency, one has 
\begin{equation}
  \label{eq:sol:tot:fin}
  \mathbf{x}_{SS} = \sum_{n=-\infty}^{+\infty} P_n\, 
  \left( U\; \frac{1}{i\,n\,\Omega - A_D}\; 
    U^{\dag} \right) \; \mathbf{w}\; \e^{i\,n\,\Omega t}.
\end{equation}

We are  interested in the $z$  component of the  magnetization, so that
after some straightforward algebra we find
\begin{equation}
  \label{eq:Mz:fin}
  \begin{split}
  M_z^{SS}(t) \propto & \;\; \Re( P_0 C_0 ) \\
  & + \sum_{n=1}^{+\infty} 
  \bigg[ 
  \Re( P_n C_n + P_{-n}C_{-n}) \, \cos n \Omega t \\
  & 
\phantom{++\sum_{n=1}^{+\infty}}
-  \Im( P_n C_n - P_{-n}C_{-n}) \, \sin n \Omega t 
\bigg]\\
\equiv & \;\; a_0 
+ \sum_n \big[ a_n \cos n\Omega t + b_n \sin n\Omega t \big],
\end{split}
\end{equation}
where 
\begin{equation}
  \label{eq:Cn:def}
  \begin{split}
    C_n =& \frac{\sin^2 \theta}{2}\left( 
    \frac{1}{i\,n\,\Omega + \gamma + i \omega_L} +
    \frac{1}{i\,n\,\Omega + \gamma - i \omega_L} \right) \\
  & + \frac{\cos^2 \theta}{i\,n\,\Omega + \gamma} .
\end{split}
\end{equation}

In  the experiment, the lock-in amplifier  detects the amplitude  of  the first  harmonic
$(n=1)$ so we  have to evaluate the term  $\sqrt{a_1^2 + b_1^2}$.  The
coefficients   $P_n$  satisfy   $P_{-n}   =  P_n^*$   for  each   $n$.
Additionally, for  odd values of $n$  we have $P_{-n} =  P_n^*= - P_n$
meaning that for  $n=1$ we can assume  $P_1 = i R_1$ and  $P_{-1} = -i
R_1$     ($R_1$     is     a     real     quantity     reported     in
Appendix~\ref{sec:der:pumping}).

Using the  condition $\Omega  \approx \omega_L$ and  $\theta=\pi/2, \,
\phi=0$ (given by the experimental conditions), after some algebra one
finds
\begin{equation}
  \label{eq:amplitude:1h}
  \mathcal{A}_1  \equiv  \sqrt{a_1^2 +  b_1^2}  = \frac{1}{\gamma^2  +
    (\Omega - \omega_L)^2 } \; |R_1| .
\end{equation}

Eq.~\eqref{eq:amplitude:1h} has a clear physical meaning: at low laser
power the  response of the system  is factored out.   The first factor
gives  the  usual resonant  behaviour  when  the modulation  frequency
$\Omega$ is swept over the magnetic resonance line.
The  second term  $R_1$ contains  the details  of the  laser frequency
modulation and level structure of the $D_1$ lines.

The  optical frequency  of the  MB  is sinusoidally  modulated at  the
magnetic resonance frequency, so that $\Omega \simeq \omega_L$ and the
laser detuning  $\delta$ from the $D_1$  $F_g=3 \longrightarrow F_e=4$
transition (see also Fig. \ref{fig:levels}) is

\begin{equation}
 \label{eq:definiz:delta}
 \delta(t)= \delta_0+\Delta\sin \omega_L t.
\end{equation}
It follows that  $R_1$ is a function of  both $\delta_0$ and $\Delta$.
Moreover  it  depends  also  on  the width  of  the  $D_1$  one-photon
transition $G=\Gamma/2  + \Gamma_c + \Gamma_{D}$,  where $1/\Gamma$ is
the  radiative lifetime  of  the excited  $D_1$ multiplet,  $\Gamma_c$
represents the  broadening due to  collisions and $\Gamma_{D}$  is the
Doppler broadening.   Due to the  presence of buffer gas,  the excited
$D_1$    states   get    depolarized   with    an    additional   rate
$\Gamma_c^{\prime}$, which  we added as  a phenomenological dependence
in $R_1$ in  a normalized form $r=\Gamma_c^{\prime}/\Gamma$.  Finally,
to model the  influence of the DB, a  parameter $\alpha$, describing a
global population imbalance of the two ground hyperfine states is also
introduced.

Appendix~\ref{sec:der:pumping}   contains   a   full  derivation   and
discussion about the explicit form of $R_1$, as well as a detailed definition of the parameter $\alpha$.

\section{Results}
\label{sec:res}

In this  section we report   experimental measurements  obtained in
different  regimes,   and  compare  them  with  the
theoretical profiles. 

Beside  atomic  constants,   the  model  contains  several  parameters
($\delta_0$,  $\Delta$,  $r$,  and  $G$)  fixed  by  the  experimental
conditions,  and  only  one   quantity,  $\alpha$,  which  is  a  free
parameter.
In our conditions  $\Gamma \approx 5$ MHz,
and  the  broadening  due  to  collisions  is  dominant, as
$\Gamma_c \approx  500$ MHz at 90  Torr of He,  and $\Gamma_{D} \approx
200$MHz, thus we use $G=0.5$ GHz in almost all the simulations.

Concerning       $r$,      it       is       known      since       the
Sixties \cite{Franzen1966,Baylis1979} 
that  the  collisions with  the  buffer  gas  atoms are  effective  in
depolarizing the $D_2$ excited  states, while perturb weakly the $D_1$
$^2P_{1/2}$ states.   Moreover our  theoretical results do  not depend
strongly on the value of $r$, and we have assumed $r = 0.5$ in all the
the simulations.
  
The only free parameter  -- $\alpha$ --  is chosen to
obtain the  best  correspondence  between  the measured  and  the  simulated
signals.  As shown  below, a  value of
$\alpha \approx  0.25 $  leads to a good comparison, a clear indication that, in spite of its very low power, DB has a  not negligible
influence.

As for  the modulation  amplitude $\Delta$, it  has to be  compared to
$\Delta_g$, and three regimes  can be identified: small, i.e. $2\Delta
\ll  \Delta_g$, intermediate  ($2\Delta \approx  \Delta_g$)  and large
($2\Delta  \gg \Delta_g$) modulation  amplitude respectively.   In the
following we discuss these three regimes.

\begin{figure}[hbtp]
\begin{center}
\includegraphics[width=.7\columnwidth]{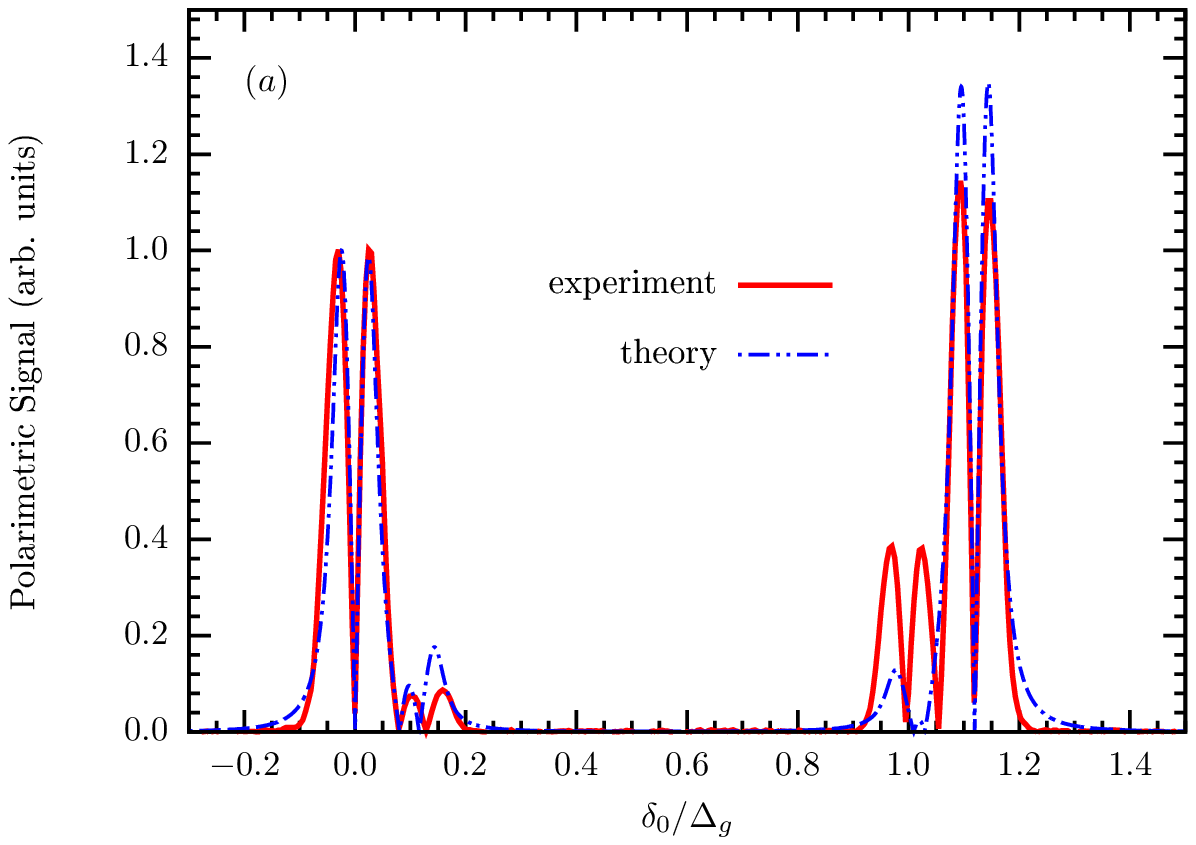}
\includegraphics[width=.7\columnwidth]{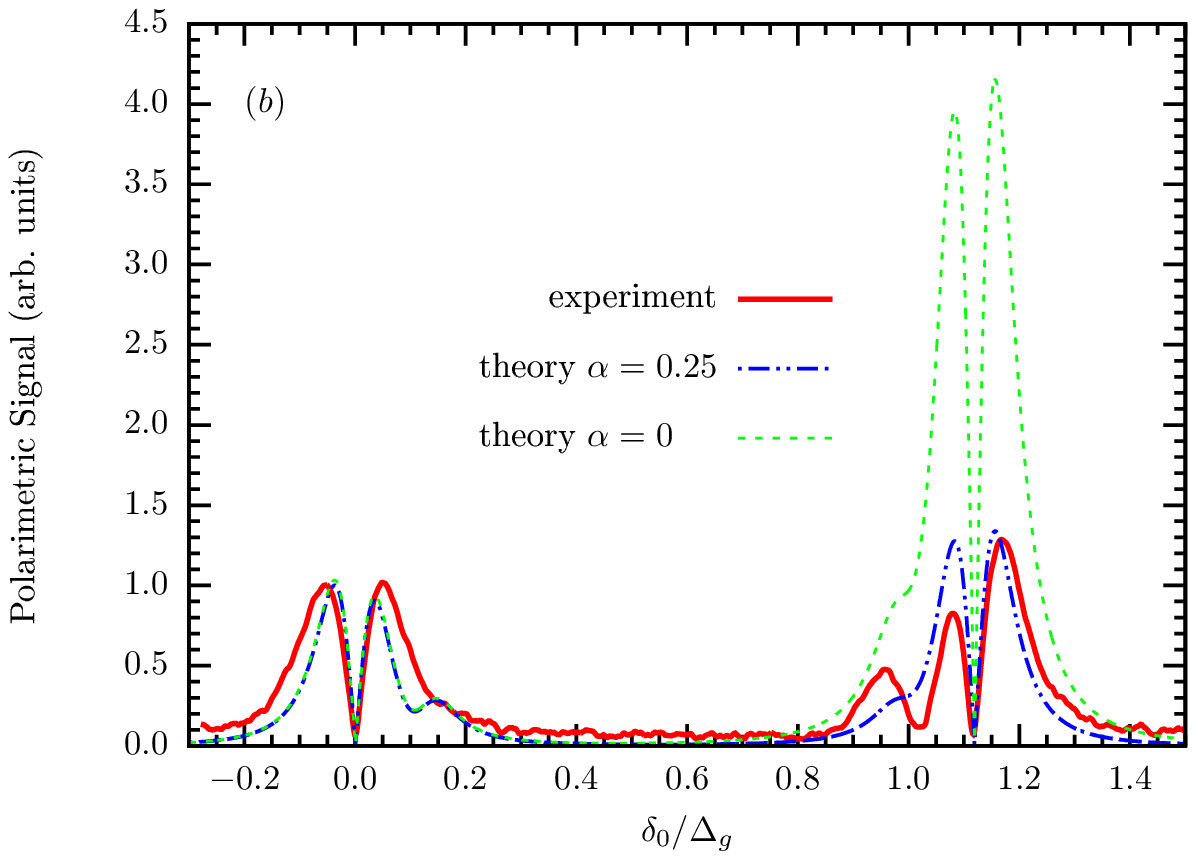}
\end{center}
\caption{
  (Color  online) Comparison  of  theoretical and  experimental
  signals  as a function  of $\delta_0/\Delta_g$ in a regime of small modulation.  For both  plots the
  following values have been
  used in the model: $2\Delta=0.5$ GHz, 
  $r=0.5$, 
  $\alpha = 0.25$, $\Delta_e = 1.1$ GHz and $\Delta_g=9.2$ GHz. 
  The plots  show magnetic resonances amplitudes as  obtained with (a)
  low buffer gas pressure (2 Torr Ar) and (b) high buffer gas pressure
  (90 Torr Ne).   Correspondingly $G = 200$ MHz and $G  = 500$ MHz are
  used in  the simulations. In the  plot (b) we report for comparison also the model output
  obtained with $\alpha=0$. 
}
\label{fig:Delta=0.5}
\end{figure}

Figure~\ref{fig:Delta=0.5} shows the signal obtained for $\Delta=0.5$
GHz.
As  predicted by Eq.\eqref{eq:forma:fin:pompaggio}, the
four $D_1$ transitions give eight peaks in $R_1$, 
separated  in two  groups around  the positions  of the  two hyperfine
ground-states, corresponding  to $\delta_0 /\Delta_g \approx  0$ and $
\delta_0  /\Delta_g \approx  1$.   These peaks  are  well resolved  in
conditions  of  small collisional  broadening  as  can  be noticed  in
Fig.~\ref{fig:Delta=0.5}  (a).    Here  the  experimental   signal  is
recorded with a  lower buffer gas pressure giving  a nominal $\Gamma_c
\approx 18$ MHz, so to compare we used the value $G = 200 $ MHz.
Increasing  the  collisional broadening  up  to  0.5  GHz, some  peaks
overlap as can be seen in the Fig.~\ref{fig:Delta=0.5} (b). 

In all the plots, we normalize to 1 the height  of the leftmost peaks, both measured and
simulated. The value  of $\alpha$ is chosen in such a  way to reproduce rightmost peaks
height  matching the  experimental observation.  With  $\alpha =  0$  the right  peak
results four  times higher than  the first one  (see the green-dashed  line in
Fig.~\ref{fig:Delta=0.5} (b)). A good  accordance between  the measured  and simulated
resonance amplitudes is found for $\alpha \approx 0.25$.

It  is remarkable that  when  the MB  is mainly  resonant with  the
$F_g=3$ transitions (e.g. $\delta_0 \approx - \Delta$) the recorded  signal has peak value 
comparable with the one  obtained with ($\delta_0 \approx \Delta_g$), in
spite of the  fact that the measured quantity  is the magnetization in
the  $F_g=4$ ground  state.   At $\delta_0  \approx  -\Delta$, the  MB
causes a  strong hyperfine pumping towards the  $F_g=4$ state. Thus, despite
the  fact  that  the  laser  is  not in  resonance  with  the  $F_g=4$
sublevels,  a high  degree of  Zeeman  pumping is  observed. Thus  the
leftmost  peak appearing  in the  plot corresponds  to  an interaction
condition where the MB  produces a high amplitude magnetic resonance,
while  weakly   perturbing  the  hyperfine  ground   state  where  the
magnetization   is  induced.    This  interaction   regime   has  been
successfully used (in a regime of stronger MB intensity) for high sensitivity
magnetometry \cite{bevilacqua_apbarx_16}.
%

\begin{figure}[hbtp]
\begin{center}
\includegraphics[width=\columnwidth]{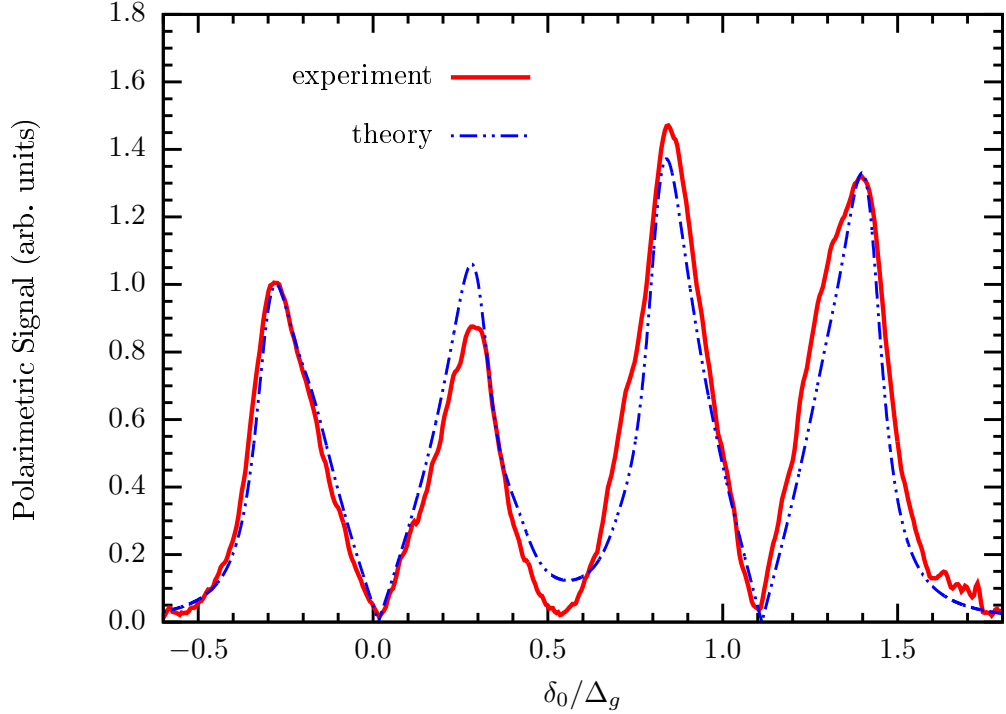}
\end{center}
\caption{(Color  online) Comparison  of  theoretical and  experimental
  signals as a function of  $\delta_0$ in the intermediate regime. The
  following values have  been used: $G = 0.5$  GHz, $2\Delta=5.6$ GHz,
  $\Gamma_c^{\prime}/\Gamma = 0.5$, $\alpha  = 0.25$, $\Delta_e = 1.1$
  GHz and $\Delta_g=9.2$ GHz. }
\label{fig:Delta=5.6}
\end{figure}

As  shown in Fig.\ref{fig:Delta=5.6},  the  model reproduces
with good  accurateness the signal  behavior also in  the intermediate
regime where $2\Delta \approx \Delta_g$.  In this case, the  MB may
resonantly  excite  either one  or both the  ground states
simultaneously,  which happens  for  $\delta_0/\Delta_g \approx  1/2$.
A good agreement  between the theoretical  and experimental
results is obtained keeping the same values of the parameters.
In  this case the eight  components merge into  four peaks of
comparable height and nearly symmetric shape.

\begin{figure}[hbtp]
\begin{center}
\includegraphics[width=\columnwidth]{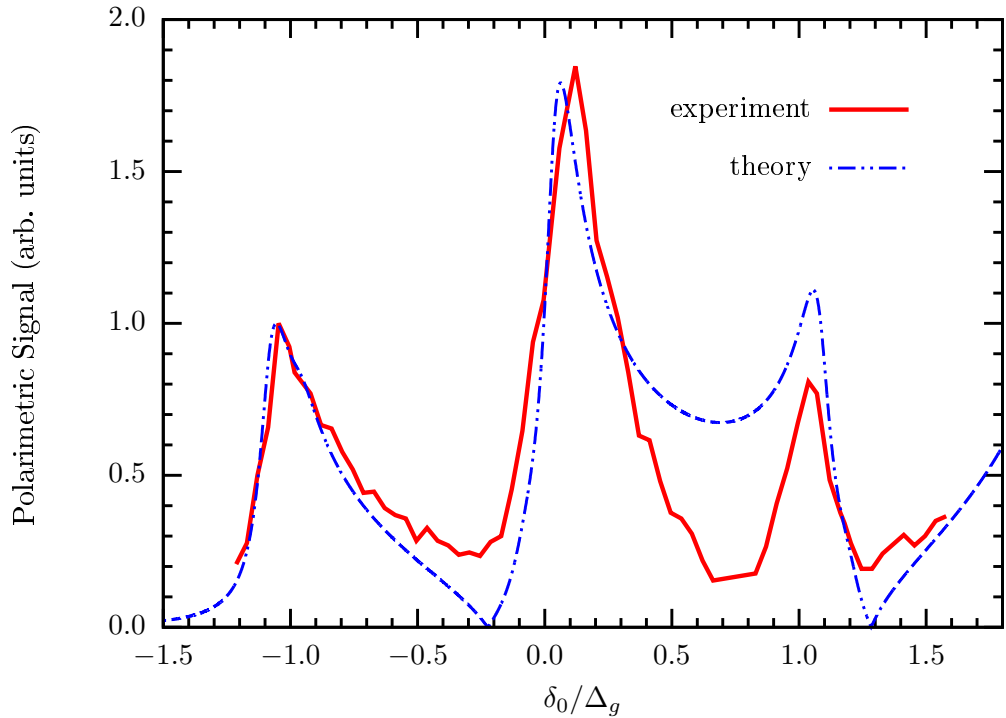}
\end{center}
\caption{(Color  online) Comparison  between  theoretical and  experimental
  signals as a function of  $\delta_0$ in the regime with $2\Delta \gg \Delta_g$. The following values have been used: $G = 0.5$  GHz, $2\Delta=20.0$ GHz, $\Gamma_c^{\prime}/\Gamma =
  0.5$, $\alpha = 0.20$, $\Delta_e = 1.1$ GHz and   $\Delta_g=9.2$ GHz. }
\label{fig:Delta=20.0}
\end{figure}

The results corresponding to the third  regime, where $2 \Delta$
exceeds   $\Delta_g$,    are  shown   in
Fig.~\ref{fig:Delta=20.0}. Here, some  technical limitations prevent the
possibility to extend the scan at higher values of $\delta_0$, so that
a     rightmost     peak     corresponding    to     $L_{eg}$     (see
Eq.~\eqref{eq:def:LeD:rightmost})  is not recorded.  The leftmost peak
has a  maximum at  $\delta_0 \approx -\Delta$,  according to  what is
expected from  Eq.\eqref{eq:def:LeD}.   The peaks observed experimentally have  an asymmetric
shape  more  evident  at  large  values of  $\Delta$,  this feature  is  well
reproduced by the model. On  the other hand, similarly to what appears
in Fig.~\ref{fig:Delta=0.5} some  discrepancies emerges more visibly at
$\delta_0  \approx \Delta_g$.  There is experimental evidence  that the  DB, in
spite  of its  very  weak  intensity, is  responsible  for these  minor
deviations: those  discrepancies actually change with the intensity and the detuning of DB.

\section{Conclusion}
A model is  developed to describe the behavior  of magnetic resonances
measured  in Cesium vapour in an  experiment  where a  weak intensity  laser
radiation  tuned to  the  D$_1$ transitions    is broadly  frequency
modulated. Such modulation makes the laser-atom interaction occur  in a condition where  both the hyperfine  ground levels are
excited. In the approximation of weak intensity, a multipole expansion
analysis enables an accurate  evaluation of the measured quantity that
is the time  dependent magnetization of atoms in  the $F_g=4$ state. A
comparison with  the experiment  is made in  three regimes,  where the
modulation  depth is  smaller, comparable  or larger  than  the ground
state  hyperfine  splitting, respectively.  A  good correspondence  is
found, and  the model reproduces satisfactorily  the recorded features
with the requirement  of  tuning   only  one  free  parameter ($\alpha$). This parameter is
phenomenologically introduced to account for an imbalance in the populations of the $F_g=3$ and $F_g=4$ states that is induced by
the detection radiation.

\appendix

\section{Derivation of the pumping term}
\label{sec:der:pumping}

Rewriting    the    Eq.~\eqref{eq:bloch}    for    each    block    of
$\boldsymbol{\rho}$       and       assuming       the       adiabatic
approximation \cite{stenholm} for  the optical coherences, for
instance we find 
\begin{equation}
  \label{eq:opt:coh}
  \rho_{e_4, g_4} = \frac{i}{G + i(\delta - \Delta_g)} 
  \; \big[ \rho_{e4,e4} \;W^{\dag}_{e4,g4} - W^{\dag}_{e4,g4} \; \rho_{g4,g4} \big]
\end{equation}
and similar expressions  for the other optical coherences  which we do
not report explicitly.  In \eqref{eq:opt:coh} $G$ is the width of the $D_1$
one-photon transition  determined as $G = \Gamma/2  + \Gamma_c $ ($1/\Gamma$  is the
lifetime  of the  excited  $D_1$ multiplet  and $\Gamma_c$  represents
additional broadening due to collisions). Finally $\delta$ is the laser
detuning  from  the  $F_g=3  \rightarrow F_e=4$  transition  (see  also
Fig.~\ref{fig:levels}). 

Substituting the expressions  like \eqref{eq:opt:coh} in the equations
for the diagonal blocks of $\boldsymbol{\rho}$ we find
\begin{subequations}
\label{eq:rho:ecc:ground}
\begin{equation}
  \label{eq:rho:ecc}
  \begin{split}
 \dot{\rho}_{e4} = & -\Gamma \rho_{e4} + {\cal L}_{coll} ( \rho_{e4} )\\
 & -i [D_0 W^{\dag}_{e4, g3}W_{g3, e4} 
+ D_g W^{\dag}_{e4, g4}W_{g4, e4}, \rho_{e4} 
] \\
& - 
 \{ L_0 W^{\dag}_{e4, g3}W_{g3, e4} 
+ L_g W^{\dag}_{e4, g4}W_{g4, e4}, \rho_{e4} 
\} \\
& + 2 L_0 W^{\dag}_{e4, g3}\; \rho_{g3} \; W_{g3, e4} \\
& + 2 L_g W^{\dag}_{e4, g4}\; \rho_{g4} \; W_{g4, e4}, 
\end{split} 
\end{equation}
where ${\cal L}_{coll}$ takes into account the collision effects in the
excited  state.  We assume  that  ${\cal  L}_{coll}$  is diagonal  and
quenches the multipoles with $k \geq 1$ (see below).

Similarly one obtains
\begin{equation}
  \label{eq:rho:ground}
  \begin{split}
    \dot{\rho}_{g4} = & -\gamma \rho_{g4} 
    - i [-\boldsymbol{\mu}\cdot\mathbf{B}, \rho_{g4} ]\\
    & -i [D_g W_{g4, e4} W^{\dag}_{e4, g4} 
    + D_{eg} W_{g4, e3} W^{\dag}_{e3, g4}, \rho_{g4} 
    ] \\
    & - 
    \{ L_g W_{g4, e4} W^{\dag}_{e4, g4} 
    + L_{eg} W_{g4, e3} W^{\dag}_{e3, g4}, \rho_{g4} \}\\
    & + 2 L_g W_{g4, e4} \; \rho_{e4} \; W^{\dag}_{e4, g4} \\
    & + 2 L_{eg} W_{g4, e3} \; \rho_{e3} \; W^{\dag}_{e3, g4} \\
    & + \mathcal{R}_{s.e}.
  \end{split} 
\end{equation}
\end{subequations}

Analogous expressions are obtained for the other diagonal blocks of
$\boldsymbol{\rho}$. From Eq.~\eqref{eq:rho:ecc:ground} we can infer that the
laser gives a Hamiltonian  contribution (term with the commutator) as
well as a  relaxation (term with the anti-commutator)  to the dynamics
of the excited and  ground states multiplets. In \eqref{eq:rho:ecc:ground} we
have introduced the abbreviations
\begin{subequations} 
  \label{eq:def:LeD}
  \begin{align}
    \frac{1}{G + i \delta} & = L_0 - i D_0, \\
    \frac{1}{G + i (\delta - \Delta_e) } & = L_e - i D_e, \\
    \frac{1}{ G + i (\delta -\Delta_g) } & = L_g - i D_g, \\
    \frac{1}{ G + i(\delta - \Delta_g - \Delta_e)} & = L_{eg} - i D_{eg} \label{eq:def:LeD:rightmost},
  \end{align}
\end{subequations}
and   $  \mathcal{R}_{s.e.}$   represents  the   spontaneous  emission
contributions,  whose  explicit  expressions  in term  of  irreducible
components  (see  below) are  reported  by  Dumont \cite{dumont}.   In
addition, we  neglect the  excited state dynamics  due to  the magnetic
field and added a phenomenological relaxation constant $\gamma$ in the
ground state.

To proceed further we assume  the low laser power limit and completely
un-polarized ground states 
\begin{subequations}
  \label{eq:assump:low:power}
  \begin{align}
    \label{eq:svil:eta}
    W \rightarrow & \; \eta W \\
    W^{\dag} \rightarrow & \; \eta W^{\dag} \\
    \rho_{e4} = & \; \eta^2 \rho_{e4}^{(2)} + O(\eta^4) \\
    \rho_{e3} = & \; \eta^2 \rho_{e3}^{(2)} + O(\eta^4) \\
    \rho_{g4} = & \; \left(\frac{1}{2}- \alpha\right) \frac{\Pi_{g4}}{2F_{g4} + 1} + \eta^2
    \rho_{g4}^{(2)} + O(\eta^4) \\
    \rho_{g3} = & \;\left(\frac{1}{2} + \alpha\right) \frac{\Pi_{g3}}{2F_{g3} + 1} + \eta^2
    \rho_{g3}^{(2)} + O(\eta^4) ,
  \end{align}
\end{subequations}
where $\eta$ is a  very small parameter quantifying the approximation.
Here  the factors  $ 1/2  \pm \alpha$  ($-1/2 \leq  \alpha  \leq 1/2$)
account, in  a phenomenological  way, for the  pumping effects  of the
DB. When $\alpha =  0 $ the DB is an ideal  probe laser not disturbing
the ground  state dynamics.  A  positive value of $\alpha$  denotes an
increase  of the  $F_g=3$  global  population and  a  decrease of  the
$F_g=4$ one.  A negative value of $\alpha$ would describe the other way around.
Introducing the  populations  imbalance  in  such simplified way corresponds to
neglect the Zeeman  sublevels structure of the ground  states and the
details of their interaction with the  DB: in other word $\alpha \neq 0$
reproduces only a  global population imbalance between the two hyperfine ground states, while excluding any 
polarization effect. 

To proceed it is better to introduce the irreducible
components \cite{omont_book_77,happer_rmp_72}  of  each density  matrix
block
\begin{equation}
  \label{eq:irreps:rho}
  \rho_{g4}^{(2)}    =     \sum_{k=0}^{2    F_{g4}}    \sum_{q=-k}^{k}
  (\rho_{g4}^{(2)})_{k,q} \; T_{k,q}(g4) ,
\end{equation}
where the irreducible tensor operators 
\begin{equation}
  \label{eq:def:Tkq}
  \begin{split}
    T_{k,q}(g4) = & \sqrt{2 k + 1 } 
    \sum_M (-1)^{F_{g4} - M } 
  \begin{pmatrix} 
    F_{g4} & F_{g4} & k \\
    M & q-M & -q 
\end{pmatrix} \\
& \times
| F_{g4} M \rangle \langle F_{g4} M - q |
\end{split}
\end{equation}
are expressed  using the  Wigner 3j coefficients.  Similar expressions
can be written for the remaining blocks. 

The effect of collisional damping in the excited state is modeled as 
\begin{equation}
  \label{eq:form:of:Lcoll}
  (   {\cal  L}_{coll}   (\rho_{e4}^{(2)})  )_{k,q}   =   -  \Gamma_c^{\prime}
  \;(\rho_{e4}^{(2)})_{k,q} \qquad k\geq 1. 
\end{equation}

The  ground  state  feeding   by  spontaneous  emission  described  by
$\mathcal{R}_{s.e.}$ in Eq.~\eqref{eq:rho:ground} assumes a simple form
for the irreducible components \cite{dumont}
\begin{equation}
  \label{eq:xi:k}
  \big[\mathcal{R}_{s.e.}(e    \rightarrow   g)    \big]_{k,q}   =
  \xi_k(J_e, Fe, J_g, Fg) \; (\rho_{e})_{k,q} ,
\end{equation}
where 
\begin{equation}
  \label{eq:def:xi:k}
  \begin{split}
    \xi_k(J_e,         F_e,        J_g,        F_g)         =        &
    (2J_e+1)(2F_g+1)(2F_e+1)\\ 
    & (-1)^{F_e+F_g+k+1} \Gamma \\
    & \SeiJ{F_e}{F_g}{1}{J_g}{J_e}{I}^2 
    \SeiJ{F_g}{F_g}{k}{F_e}{F_e}{1}. 
  \end{split}
\end{equation}
After some algebra Eq.~\eqref{eq:rho:ground} becomes 
\begin{equation}
  \label{eq:rho:ground:WW}
  \begin{split}
    \frac{\de}{\de  t}  (\rho_{g4}^{(2)})_{k,q}\bigg|_{LASER}  = &  \;
    \left( \frac{1}{2}-\alpha \right) \frac{L_g}{9} \bigg[ -( W_{g4,e4}
    W_{e4,g4}^{\dag})_{k,q}         +        \frac{\xi_k(e4\rightarrow
      g4)}{\Gamma_c^{\prime}}
    ( W_{e4,g4}^{\dag}W_{g4,e4})_{k,q} \bigg] + \\
    &  \;\left(\frac{1}{2}-\alpha  \right)  \frac{L_{eg}}{9} \bigg[  -(
      W_{g4,e3}  W_{e3,g4}^{\dag})_{k,q}  +  \frac{\xi_k(e3\rightarrow
        g4)}{\Gamma_c^{\prime}}
      ( W_{e3,g4}^{\dag}W_{g4,e3})_{k,q} \bigg] + \\
      & \; \left(\frac{1}{2}+\alpha \right) \frac{1}{7}
      \bigg[     L_0    \frac{\xi_k(e4\rightarrow
        g4)}{\Gamma_c^{\prime}}  (  W_{e4,g3}^{\dag}W_{g3,e4})_{k,q} +
      L_e    \frac{\xi_k(e3\rightarrow    g4)}{\Gamma_c^{\prime}}    (
      W_{e3,g3}^{\dag}W_{g3,e3})_{k,q} \bigg].
\end{split}
\end{equation}
Using  standard  methods  (see \cite{omont_book_77})  the  irreducible
components of $W \,W^{\dag}$ and $W^{\dag}W$ can be worked out
\begin{subequations}
  \label{eq:WW:decomp}
\begin{align}
  \label{eq:WWdag:decomp}
    (W_{g_i, e_j}W_{e_j, g_i}^{\dag})_{k,q} & = E_0^2 \langle
    F_{e_j}||\mathbf{d} || F_{g_i}\rangle ^2 (-1)^{F_{e_j} - F_{g_i}} 
    \SeiJ{1}{1}{k}{F_{g_i}}{F_{g_i}}{F_{e_j}}\,                  (-1)^q
    \mathbb{E}_{k,-q}\\
    (W_{e_j, g_i}^{\dag}W_{g_i, e_j})_{k,q} & = E_0^2 \langle
    F_{e_j}||\mathbf{d} || F_{g_i}\rangle ^2 (-1)^{ F_{g_i} - F_{e_j}} 
    \SeiJ{1}{1}{k}{F_{e_j}}{F_{e_j}}{F_{g_i}}\,                  (-1)^q
    \mathbb{E}_{k,-q}.
\end{align}
\end{subequations}

The reduced  matrix element  of the dipole  can be  rewritten as \cite{judd}
\begin{equation}
  \label{eq:dipole:red}
  \begin{split}  
    \langle F_{e_j}||\mathbf{d} || F_{g_i}\rangle & \equiv
    \langle (J_e I) F_{e_j}||\mathbf{d} || (J_g I) F_{g_i}\rangle \\
    & = (-1)^{J_e+I+F_{g_i}+1}\sqrt{(2F_{e_j}+1)(2F_{g_i}+1)} 
    \SeiJ{F_{e_j}}{1}{F_{g_i}}{J_g}{I}{J_e} 
\langle  J_e || \mathbf{d}    || J_g \rangle,
  \end{split}
\end{equation}
while the  polarization tensor $\mathbb{E}_{k,q}$  is constructed from
the laser polarization vector as
\begin{equation}
  \label{eq:E:kq}
  \mathbb{E}_{K,Q} = (-1)^{K+Q}\sqrt{2K+1} 
  \sum_{q,q^{\prime}=-1}^{1}\TreJ{1}{1}{K}{q}{q^{\prime}}{Q}\; 
  (\epsilon^*)_{-q}\, \epsilon_{-q^{\prime}}, 
\end{equation}
which for circular $\sigma^+$ polarization becomes
\begin{equation}
  \label{eq:Ekq:sigmap}
  \mathbb{E}_{k,q} = - \delta_{q,0}
  \left( 
    \frac{1}{\sqrt{3}}\delta_{k,0} + 
    \frac{1}{\sqrt{2}}\delta_{k,1} +
    \frac{1}{\sqrt{6}}\delta_{k,2} 
  \right) .
\end{equation}

Putting all together Eq.~\eqref{eq:rho:ground:WW} becomes 
\begin{equation}
  \label{eq:forma:fin:pompaggio}
  \begin{split}
    \frac{\de}{\de t} (\rho_{g4}^{(2)})_{1,q}\bigg|_{LASER} = & 
    -\frac{\sqrt{15}}{20736}  E_0^2 \langle J_e  || \mathbf{d}  || J_g
    \rangle^2 \frac{1}{1+r} \times  \\
    & \bigg[ 
    (1-2\alpha)(29+48r)\,L_g \\ 
    & \phantom{+} + 
    21(1-2\alpha)(25+16 r) L_{eg} \\
    & \phantom{+} - 171(1+2\alpha) L_0
    - 27 (1+2\alpha) L_e
    \bigg]\delta_{q,0} , 
  \end{split}
\end{equation}
where $r=\Gamma_c^{\prime}/\Gamma$.  Dropping the constant (irrelevant
at this  order of approximation) in  front of the  expression, this is
exactly  the $P(t)$  function used  in Eq.~\eqref{eq:fourier:pumping}.
The  time-dependence  arises  from  the  laser  modulation,  i.e.,  in
Eq.~\eqref{eq:def:LeD} the substitution $\delta \rightarrow \delta_0 +
\Delta\sin \Omega t$.

The Fourier coefficients  $P_n$ of Eq.~\eqref{eq:fourier:pumping} have
an analytical  form. In fact  re-doing the steps  of \cite{lorenz:mod}
one finds ($n\geq 0$)
\begin{equation}
  \label{eq:coeff:lor}
  \begin{split}
  P_n^{(0)} \equiv & \frac{\Omega}{2\pi} \int_0^{2\pi/\Omega}
  \e^{-i\,n\,\Omega\,t}\, L_0(t) \de t \\
  = & \frac{1}{2\pi}  \int_0^{2\pi} \e^{-i\,n\,\theta} \, \frac{G}{G^2
    + (\delta_0 + \Delta \sin\theta)^2} \de\theta \\
  = & \frac{1}{2}\int_{-\infty}^{+\infty} J_n(z\Delta) 
  \e^{i \,z \,\delta_0} \e^{-G\,|z|} \de z\\
  = & 
  \begin{cases}
    \Re( I_n ) & n \; \mathrm{even} \\
    i\Im( I_n) & n \; \mathrm{odd} ,
  \end{cases}
  \end{split}
\end{equation}
where 
\begin{equation}
  \label{eq:def:In}
  \begin{split}
  I_n \equiv & \int_{0}^{+\infty} J_n(z\Delta) 
  \e^{i \,z \,\delta_0} \e^{-G\,z} \de z \\
  =  & \frac{1}{\Delta^n}  \frac{\bigg[ \sqrt{  (G -  i  \delta_0)^2 +
      \Delta^2} - (G-i\delta_0) \bigg]^n}{\sqrt{  (G - i \delta_0)^2 +
      \Delta^2} }
\end{split}
\end{equation}
and    the   last    step    follows   from    formula   (6.611)    of
\cite{gradshteyn2007}. So the first harmonic coefficient reads as (see
also Eq.~\eqref{eq:amplitude:1h})
\begin{equation}
  \label{eq:1h:coeff}
  R_1^{(0)} = -\frac{1}{\Delta} 
  \Im \left( \frac{G-i\delta_0}{\Delta}\,
    \frac{1}{\sqrt{1+\left(\frac{G-i\delta_0}{\Delta}\right)^2} }
    \right) ,
\end{equation}
which can be rewritten using the dispersive and absorptive profiles 
\begin{subequations}
  \begin{align}
  \label{eq:D:L}
  \mathcal{D}(\delta_0) & = 
  \frac{\delta_0 -\Delta}{ (\delta_0-\Delta)^2 + G^2 } - 
  \frac{\delta_0 + \Delta}{ (\delta_0+\Delta)^2 + G^2 } \\
  \mathcal{L}(\delta_0) & = 
  \frac{1}{ (\delta_0-\Delta)^2 + G^2 } + 
  \frac{1}{ (\delta_0+\Delta)^2 + G^2 } 
\end{align}
\end{subequations}
as 
\begin{equation}
  \label{eq:fin:form:R1}
  R_1^{(0)} = -\frac{1}{\sqrt{2}\Delta}\, \mathrm{sign}(-\delta_0) 
  \left[ 
  \sqrt{1+\Delta\frac{G^2 + 3 \Delta^2/4}{G^2+\Delta^2}
    \left(
      \mathcal{D}(\delta_0)   +   \frac{\Delta^2}{4G^2  +   3\Delta^2}
      \mathcal{L}(\delta_0) 
    \right) 
  } 
  - 1 - (\Delta/2) \mathcal{D}(\delta_0)
\right]^{1/2}. 
\end{equation}

\begin{figure}[hbtp]
\begin{center}
\includegraphics[width=\columnwidth]{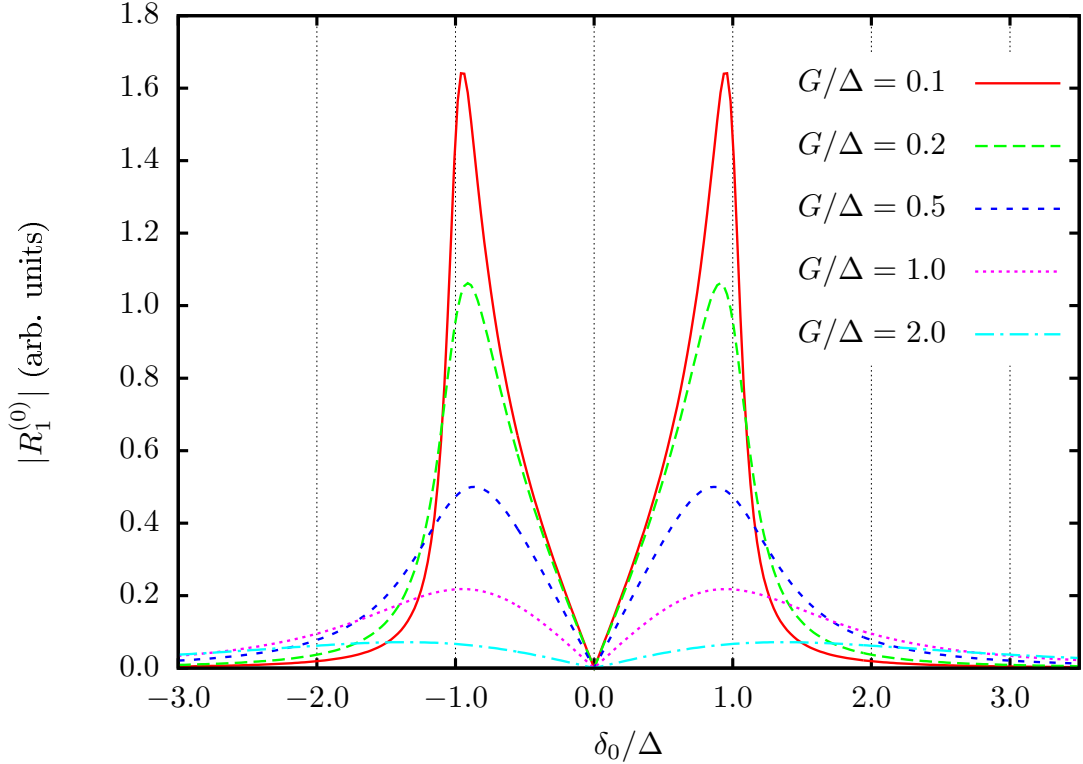}
\end{center}
\caption{(Color  online)  
Typical   profile   obtained  from   the   excitation   of  a   single
transition. The peaks are located in  correspondence of $\delta_0
\approx \pm \Delta$ for $G/\Delta \lesssim 1$. At larger values of $G$ the
peaks broaden and start shifting in opposite directions. 
}
\label{fig:prof}
\end{figure}

This is  the contribution of  $L_0(t)$ that is the  $F_g=3 \rightarrow
F_e=4$ line and it is shown
 in Fig.~\ref{fig:prof}. 

 Similar
expressions  hold for the  other transitions  and adding  all together
with the coefficients  of Eq.~\eqref{eq:forma:fin:pompaggio} we obtain
the  whole  $R_1$  which   contains  the  dependence  from  the  laser
modulation parameters.

\bibliographystyle{apsrev}\bibliography{op_bibtex}

\end{document}